\documentclass[]{aa}

\usepackage{natbib}
\usepackage{graphics}
\usepackage{graphicx}
\usepackage{epsfig}
\usepackage{amssymb}

\usepackage{epsf,psfig}
\usepackage[dvips]{color}
\usepackage[latin1]{inputenc}
\usepackage[T1]{fontenc}
\usepackage{color}
\definecolor{red}{rgb}{0.7,0,0}
\definecolor{blue}{rgb}{0,0,0.7}

\def\ki{$\chi^2$ }

\def\etal{et~al.~}

\def\kir{$\chi^2_{red}$}
\def\th{$^{\mathrm{th}}$}

\begin{document}

\title{The broad-band spectrum of Cygnus X-1 measured by INTEGRAL}
\author{M. Cadolle Bel\inst{1,2}, P. Sizun\inst{1}, A. Goldwurm\inst{1,2},
 J. Rodriguez\inst{1,3,4}, P. Laurent\inst{1,2}, A. A. Zdziarski\inst{5},
L.~Foschini\inst{6}, P. Goldoni\inst{1,2}, C.~Gouiff\`es\inst{1}, J.
Malzac\inst{7}, E.~Jourdain\inst{7}, J.~-P.~Roques\inst{7}}

\offprints{M. Cadolle Bel : mcadolle@cea.fr}

\institute{Service d'Astrophysique, CEA-Saclay, 91191, Gif-Sur-Yvette, France
\and APC-UMR 7164, 11 place M. Berthelot, 75231 Paris, France \and
AIM-UMR 7158, France \and ISDC, 16 Chemin d'Ecogia, 1290 Versoix, Switzerland \and
N. Copernicus Astronomical Center, 00-716 Warsaw, Poland \and
INAF/IASF, Sezione di Bologna, Via Gobetti 101, 40129 Bologna, Italy \and
Centre d'Etude Spatiale des Rayonnements,31028 Toulouse, France}

\date{Received ; accepted}
\authorrunning{M. Cadolle Bel et al.}
\titlerunning{High-Energy {\it INTEGRAL} Observations of Cygnus X-1}

\abstract {The {\it INTEGRAL} satellite extensively observed the
black hole binary Cygnus~X-1 from 2002 November to 2004 November
during calibration, open time and core program (Galactic Plane
Scan) observations. These data provide evidence for significant
spectral variations over the period. In the framework of the
accreting black hole phenomenology, the source was most of the
time in the Hard State and occasionally switched to the so-called
``Intermediate State''. Using the results of the analysis
performed on these data, we present and compare the spectral
properties of the source over the whole energy range
(5~keV--1~MeV) covered by the high-energy instruments on board
{\it INTEGRAL}, in both observed spectral states. Fe line and
reflection component evolution occurs with spectral
changes in the hard and soft components. The observed behaviour of
Cygnus~X-1 is consistent with the general picture of galactic
black holes. Our results give clues to the physical changes that
took place in the system (disc and corona) at almost constant
luminosity during the spectral transitions and provide new 
measures of the spectral model parameters. In
particular, during the Intermediate State of 2003 June, we observe
in the Cygnus~X-1 data a high-energy tail at several hundred keV
in excess of the thermal Comptonization model which suggests the
presence of an additional non-thermal component.

}
 \maketitle

\section{Introduction}

Galactic Black Holes (BH) X-ray binary systems display high-energy
emission characterized by spectral and flux variabilities on time
scales ranging from milliseconds to months. These systems are
generally found in two major states mainly defined by the relative
variable contributions of soft and hard X-ray components, radio
spectral properties and timing behaviour
\citep{McClintock03,nowak03}. In the Hard State (HS), the X-ray
and $\gamma$-ray spectrum is generally described by a power law
model with an exponential cutoff at a few hundred keV, accompanied
by relevant radio emission; it can be modeled by thermal
Comptonization of cool seed photons in a hot electron plasma
\citep{gier97,dove98}. The soft ($\sim$ 0.1--2 keV) black body
component is very weak or too soft to contribute significantly.
The Thermal Dominant State (TDS) instead shows a strong thermal
component with a characteristic temperature of at most a few keV
that dominates the X-ray spectrum. No, or very weak and spectrally
steep, hard X-ray emission is observed; the radio emission is
quenched or very faint. This spectrum is generally associated with
a geometrically thin and optically thick accretion disc
\citep{shak73}. In addition to these two canonical states, other
states have been identified, characterized either by an even
greater luminosity than in the TDS (the ``Steep Power law State'')
or by variability and X-ray spectral properties mostly
intermediate between the HS and the TDS
\citep{bell96}: the ``Intermediate State'' (IS).\\
\indent Cygnus~X-1/HDE 226868 is one of the first X-ray binaries
detected; it belongs to the BH binary category. Among the
brightest X-ray sources of the sky, it is also very variable on
different time scales. The assumption that Cygnus~X-1 ranks among
the microquasars has been confirmed by the detection of a
relativistic jet \citep{stir01}. Since its discovery in 1964
\citep{bow65}, it has been extensively observed as the prototype
of BH candidates in radio/optical wavelengths and with all
high-energy instruments, from soft X-rays to $\gamma$-rays, e.g.,
with {\it ASCA} \citep{gier99}, {\it SIGMA} (Salotti \etal 1992),
{\it RXTE} \citep{dove98,pott03a}, {\it BeppoSAX}
\citep{front01,disalvo01} and {\it CGRO} \citep{mc00,mc02}. This
persistent source, located at $\sim~2.4~\pm~0.5$~kpc (McClintock
\& Remillard 2003, Table 4.1), accretes via a strong stellar wind
coming from its companion, a O9.7I star of 20~M$_\odot$
\citep{zio05} with an orbital period of 5.6 days. The mass
function constrains the inclination angle of the system
between 25° and 67° \citep{gier99} and we adopted the value of 45°.\\
\indent Cygnus~X-1 spends most of its time (90$\%$ until 1998 see, e.g.,
Gierli\'nski \etal 1999) in the HS, with a relatively low
flux in soft X-rays ($\sim$~1~keV) and a high flux in hard
X-rays ($\sim$~100~keV). Its spectrum is roughly described by a
power law with a photon index $\Gamma$ between 1.4--2.1; a 
break at energies higher than $\sim$~50~keV is present. This 
state is also characterized by a large timing variability. 
Occasionally, the source switches to the TDS
with $\Gamma$~>~2.3. During 1996 June, in addition to the
dominant black body component and the hard component,
a high-energy tail extending up to 10~MeV was
discovered \citep{mc02}. In this state little variability 
is observed. The IS, in which the source exhibits a relatively 
soft hard X-ray spectrum ($\Gamma$~$\sim$~2.1--2.3) and a moderately
strong soft thermal component \citep{men97}, often appears when 
the source is about to switch from one main
state to another. When not associated with a clear spectral transition,
this state is called a ``Failed State Transition'' (FST). In the IS, the source displays a complicated pattern of timing properties. 
In the past few years, the source has
been deeply observed in the IS and in the TDS
\citep{zdziarski02,pott03a,gleiss04,zd04}. In addition to the thermal 
and power law components, other spectral features can
be present in the spectrum: a reflection component peaking around 30~keV and,
most noticeably, a Fe K$\alpha$ line and Fe edge between 6 and 7 keV.
These features can be visible in both spectral states. \\
\indent The {\it INTErnational Gamma-Ray Astrophysics Laboratory
(INTEGRAL)} mission \citep{winkler03} is an European Space Agency
satellite launched on 2002 October 17, carrying two main
 $\gamma$-ray instruments, IBIS \citep{ubertini03} and SPI \citep{vedrenne03},
and two X-ray monitors JEM-X \citep{lund03}. Composed of two
detectors, ISGRI \citep{lebrun2003} and PICsIT
\citep{dicocco03}, the IBIS coded mask instrument covers the
energy range between 15~keV and 10~MeV. The SPI telescope works in
the 20~keV--8~MeV range and the JEM-X~monitors provide spectra and
images in the nominal 3--35~keV band. As a bright hard X-ray
source, Cygnus~X-1 is a prime target for {\it INTEGRAL} and has
been extensively observed during the Performance Verification (PV)
Phase of the mission, when the source was in the HS
\citep{bazz03,bouch03,cad04a,pott03b}. Pottschmidt \etal (2005)
also reported on preliminary analysis of the high time-resolution
Galactic Plane Scan (GPS) observations of Cygnus~X-1 
(up to 2004 April) in the 15--70~keV
range. In the present work, we report the results over a wide energy
 band (from 5~keV up to 1~MeV) of several sets of observations
of Cygnus~X-1, including part of the PV-Phase
observations not yet exploited, the first observations of
Cygnus~X-1 in the Open Time program, a larger amount of
GPS data than previously analyzed and the data from the 2004 November
calibration period.
For the first time, up to 1.5 Ms of {\it INTEGRAL} data of Cygnus~X-1,
collected over two years from 2002 November to 2004 November, are
presented, fully exploiting the broad-band capability
of all high-energy instruments of the mission.

\begin{figure}
\centering\includegraphics[width=1\linewidth]{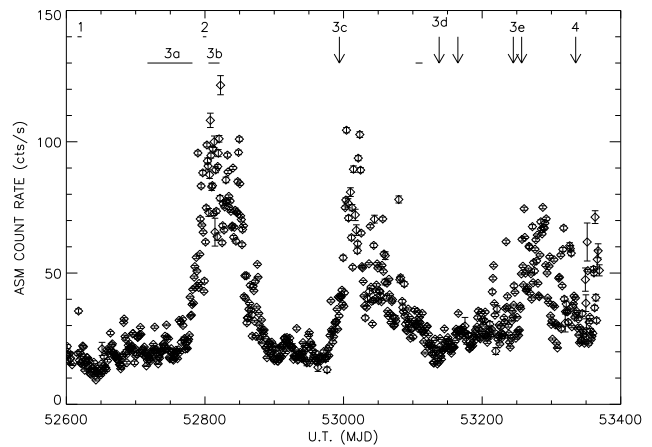}
\caption{\label{LCsimult}{\it RXTE}/ASM daily average
(1.5--12~keV) light curve of Cygnus~X-1 from 2002 November to 2004
November (MJD~=~JD~$-$~$2 400 000.5$) with the periods of our {\it
INTEGRAL} observations (see text and Tables 1 and 2 for epoch definitions).}
\end{figure}

\section{Observations and data reduction}

Table~\ref{tab:log} reports the general periods (epochs) of the
observations used, giving for each of them the
instrument data available, date range, exposure (per instrument)
and observing modes. Epoch 1 includes part of
the PV-Phase observations of Cygnus X-1. To discuss the time
evolution of the source, we report here the IBIS/ISGRI light
curves and hardness ratios obtained during most of the PV-Phase
observations of Cygnus~X-1 (first line of Table~\ref{tab:log}),
but since spectral results were presented in previous works, we
studied more specifically the broad-band spectrum (using JEM-X,
IBIS and SPI data) only for those PV-Phase observations not yet
fully exploited, i.e., those performed between 2002 December 9--11
(epoch 1). The Open Time observation was performed on 2003 June
7--11 (epoch 2) with a 5$\times$5 dither pattern \citep{jen03}.
The effective exposure time was 275~ks for JEM-X2, 292~ks for IBIS
and 296~ks for SPI. For this latter period, preliminary results
can be found in Malzac \etal (2004) and Cadolle Bel \etal (2004)
but we report here the complete study of the average spectrum
while, in a future work, Malzac \etal (submitted) will
present the variability properties of the source. Epoch 3 and
epoch 4 refer respectively to the set of Cygnus~X-1 observations
during the core program GPS and the 2004 November calibrations.
Unfortunately, for the short interrupted GPS exposures, the SPI
data are unusable because a sensitive evaluation of the background
is not possible: only JEM-X and IBIS/ISGRI data are used. The
periods of our different {\it INTEGRAL} observations, presented in
Table~\ref{tab:log}, are also indicated in Fig.~\ref{LCsimult}
(epochs from numbers 1 to 4).\\
\indent We reduced the IBIS and JEM-X data with the standard
analysis procedures of the Off-Line Scientific Analysis {\tt
OSA~4.2}~ released by the ISDC, whose algorithms are described in
Goldwurm \etal (2003) and Westergaard \etal (2003) for IBIS and
JEM-X respectively. A basic selection was performed to exclude
those pointings too close to radiation belt entry or exit, or
spoilt by too much noise. To avoid uncertainties in
the response files for high off-axis angles, we also selected the
IBIS data of the observations for which the source was in the
fully coded field of view, i.e., with an offset from the telescope
axis no larger than 4.5°, and JEM-X data for maximum offset angles
of~3°. Following recommendations of the {\tt OSA~4.2}~release,
IBIS/ISGRI events were selected to have corrected energies
$>$~20~keV and rise time channels between 7 and 80. For the
background correction, we used a set of IBIS/ISGRI maps derived in
256 energy channels from empty field observations (these maps will
be the default IBIS/ISGRI correction maps for the {\tt
OSA~5.0}~release) combined with our chosen energy bins while, for
the off-axis correction maps and the response matrices, we used
those of the official {\tt OSA~4.2}~release. In the analysis, we
considered the presence of the two other sources of the region,
Cygnus~X-3 and EXOSAT 2030$+$375, when they were active. For the
IBIS/ISGRI spectral extraction however, we implemented the most
recent module (prepared for the {\tt OSA 5.0} delivery) which is
based on the least squares fit done on background and efficiency
corrected data, using coded source zones only. This option
minimizes spurious features in the extracted spectra, which appear
in particular when the sources are weak, partially coded and the
background poorly corrected (A. Gros, private communication). 
For the PICsIT spectral extraction, we took the flux and error
values in the mosaic image at the best-fit position found for the
source. We used the response matrices officially released with
{\tt OSA~4.2},
rebinned to the 8 energy channels of the imaging output.\\
\indent The SPI data were preprocessed with {\tt OSA~4.2}
using the standard energy calibration gain
coefficients per orbit and excluding bad quality
pointings which have anomalous exposure and dead time values (or with a high
final~\ki during imaging). The {\tt spiros~9.2}~release \citep{skin03}
was used to extract the spectra of Cygnus~X-1,
Cygnus~X-3 and EXO~2030$+$375, with a background model proportional
to the saturating event count rates in the Ge detectors.
Concerning the instrumental response, version 15 of the IRF (Image
Response Files) and version 2 of the RMF (Redistribution Matrix Files)
were used for epoch 1 and 2, e.g., prior to detector losses, while
versions 17 and 4 respectively were taken for epoch 4, e.g., after
the failure of two detectors.

\begin{table*}[htbp]
\begin{center}
\caption{\label{tab:log} Log of the Cygnus~X-1 observations analyzed in this
paper.}
\begin{tabular}[h]{lllll}
\hline
Epoch &~~~Instrument & Observation Period &~Exposure & Observation\\
&&(date yy/mm/dd)&~~~~(ks) &~~~~Type \\
\hline
~~~~~ &~~~~~~~IBIS & 02/11/25--02/12/15 &~~~~~810 &staring, 5$\times$5~$^a$, hex~$^b$\\
~~~~1 & IBIS/SPI/JEM-X & 02/12/09--02/12/11 & 365/365/31 &~~~~5$\times$5 \\
~~~~2 & IBIS/SPI/JEM-X & 03/06/07--03/06/11 & 292/296/275 &~~~~5$\times$5\\
~~~~3 &~~IBIS/JEM-X & 03/03/24--04/09/10 &~~269/35 &~~~~GPS~$^c$ \\
~~~~4 & IBIS/SPI/JEM-X &~~~~~~04/11/22 &~~~8/8/6 & calibration  \\
\hline
\end{tabular}
\end{center}
Notes:\\
~a) 5$\times$5 dither pattern around the nominal target location.\\
~b) hexagonal pattern around the nominal target location.\\
~c) individual exposures separated by 6° along the scan path, shifted by 27.5°
in galactic longitude.\\
(Observations indicated on the first line also used, together with epochs 1--4, for Fig.~\ref{HR1} and~\ref{HR}.)
\end{table*}

\begin{figure}
\centering\includegraphics[width=1\linewidth]{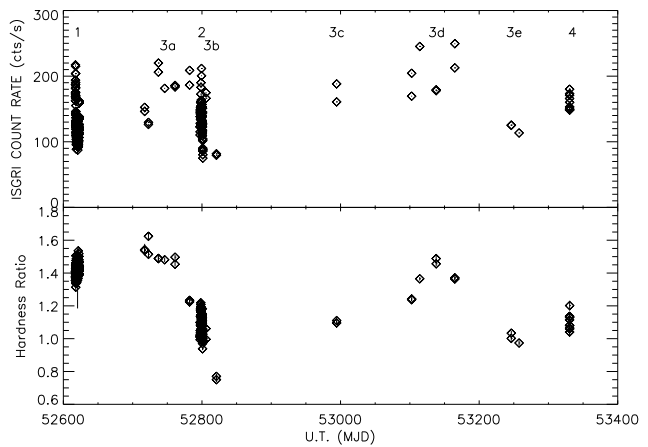}
\caption{\label{HR1}~The 20--200 keV IBIS/ISGRI light curve of
Cygnus~X-1 from 2002 November 25 until 2004 November 22 and corresponding HR
between the 40--100 and the 20--30 keV energy bands 
(see text and Tables 1 and 2 for epoch definitions).}
\end{figure}

\begin{figure}
\centering\includegraphics[width=1\linewidth]{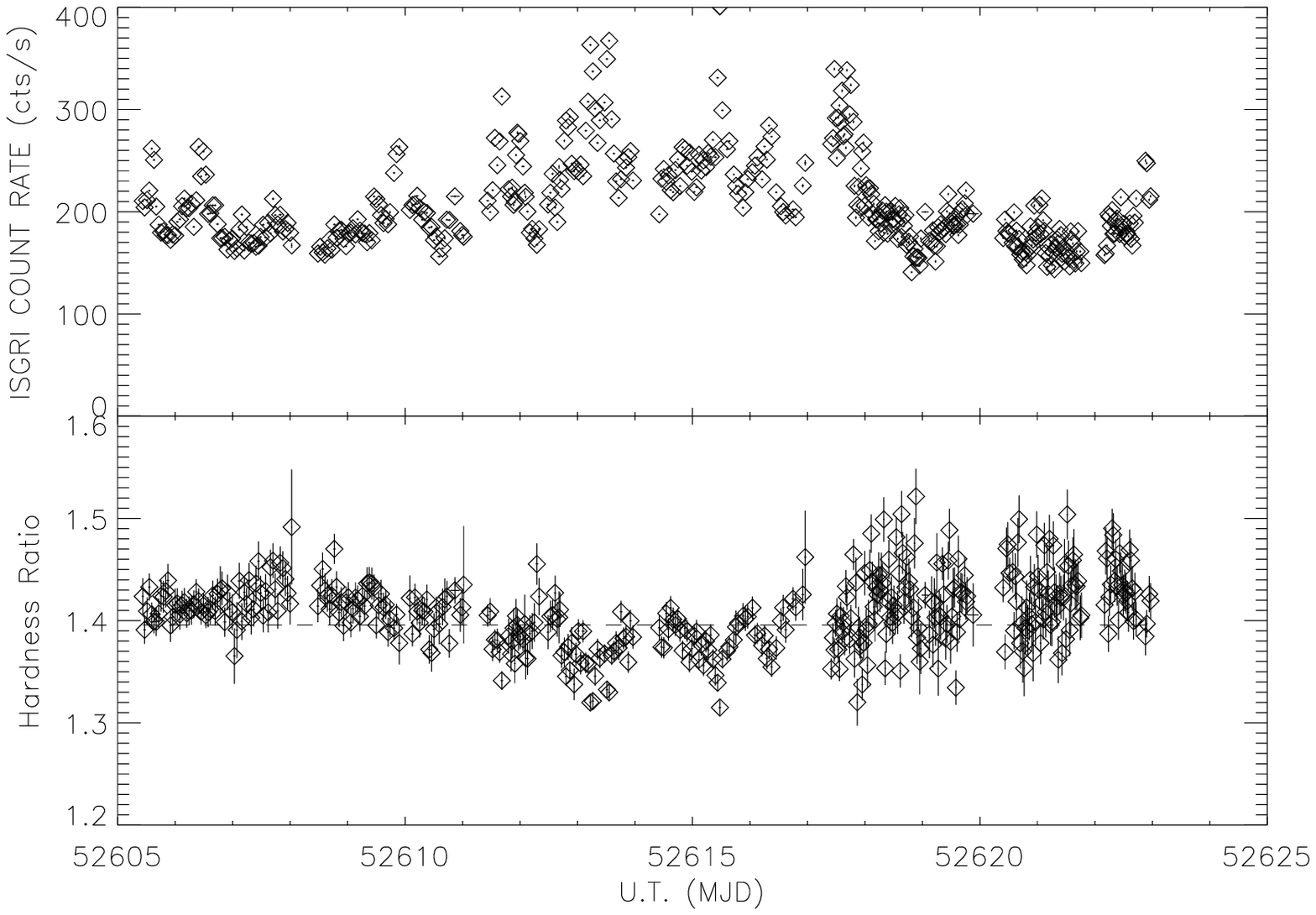}
\centering\includegraphics[width=1\linewidth]{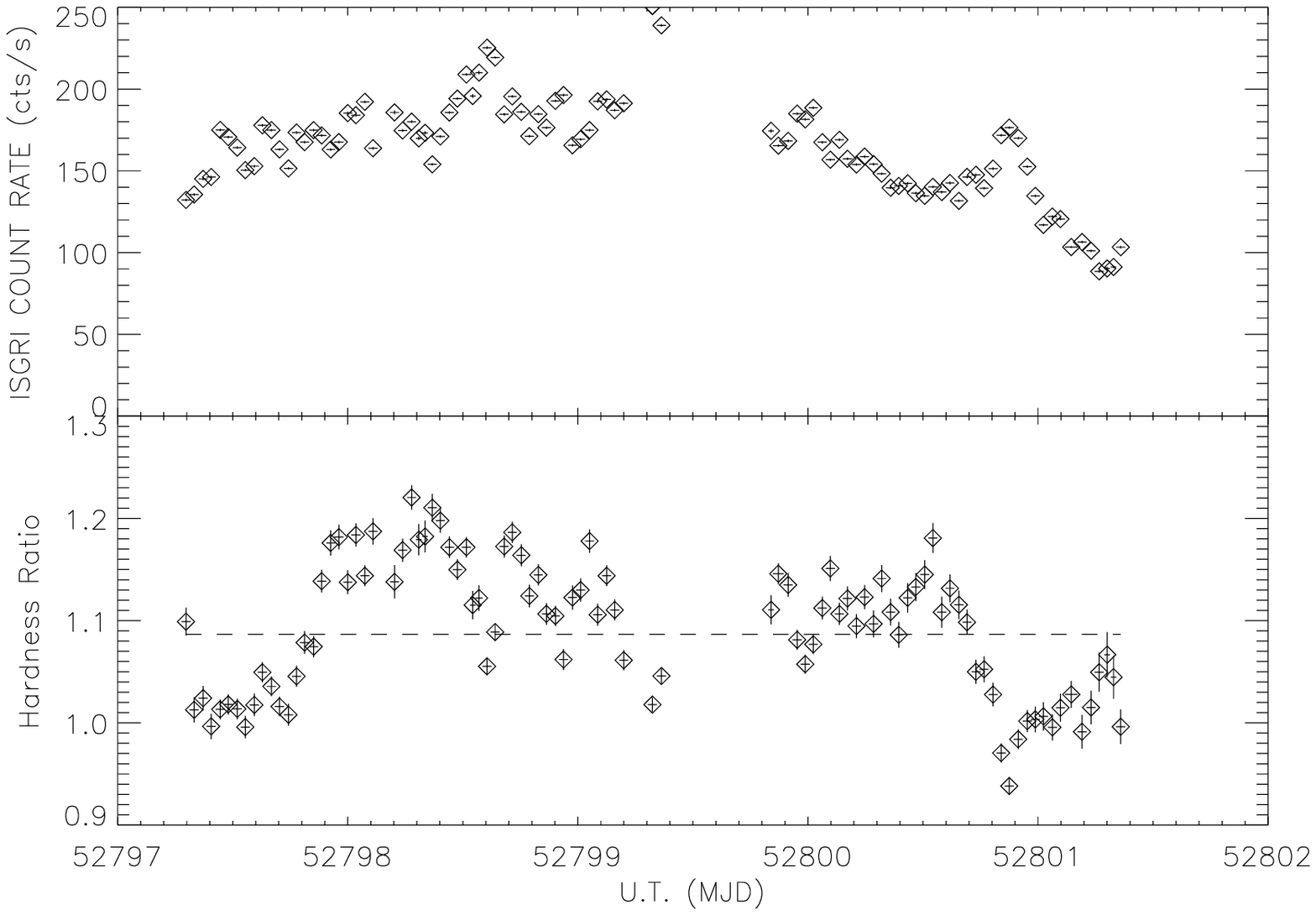}
\caption{\label{HR}~{\bf{Top:}}~Zoom on the 20--200 keV IBIS/ISGRI light curve of
Cygnus~X-1 from 2002 November 25 until December 15 and corresponding HR between
the 40--100 and the 20--30 keV energy bands (average level denoted by dashed line).
~{\bf{Bottom:}}~Same as above for epoch 2.}
\end{figure}

\section{Results of the analysis}

As shown in Fig.~\ref{LCsimult}, during the epoch 2 {\it INTEGRAL}
observations, the 1.5--12~keV ASM average count rate of Cygnus~X-1
($\sim$~1.3 Crab) was larger than during epoch 1 ($\sim$~290
mCrab) by a factor of 4.5. The derived IBIS/ISGRI 20--200 keV light
curves and Hardness Ratio (HR) of Cygnus~X-1 are shown
respectively in Fig.~\ref{HR1} (general overview of part of
PV-Phase and epochs 1 to 4) and Fig.~\ref{HR} (specific zooms on
part of PV-Phase, epochs 1 and 2). From epoch 1 to epoch 2, while
the ASM average count rate increased, 
the 20-200 keV IBIS/ISGRI one
decreased from $\sim$~910 to $\sim$~670~mCrab as shown in
Fig.~\ref{HR} (where, in the 20-200~keV range, 1~Crab =
205~cts~s$^{-1}$). This probably indicates a state transition
between epochs 1 and 2, as also suggested by the decrease in
the IBIS HR (Fig.~\ref{HR}). Similar transition, with a change in
the ASM light curves and an evolving IBIS HR, occurred again
during GPS data (epoch 3). Figure~\ref{HR1} (bottom) shows the
IBIS HR behaviour over the whole 2002--2004 period indicated in Table~\ref{tab:log}.\\
\indent We sampled epoch 3 in five distinct sub-groups 
(noted {\it{a}} to {\it{e}}) of close pointings 
which appear to occur, according to Fig.~\ref{LCsimult} and~\ref{HR1}, 
in different regimes of ASM count rate and of average IBIS HR.
The data of each epoch (and sub-group) were 
summed to obtain an average spectrum studied
separately. We added 3$\%$ systematic errors for JEM-X (5--30~keV
range), IBIS (20~keV--1~MeV range) and SPI (22~keV--1~MeV range)
and fitted the resultant spectra simultaneously using {\tt XSPEC
v11.3.0}  \citep{arnaud96}. In order to account for uncertainties
in the cross-calibration of each instrument, a multiplicative
constant was added in the spectral fits to each instrument data
set: it was set free for IBIS and SPI and
frozen to 1 for JEM-X.\\
\indent Several models were used in the course of the spectral
analysis. In {\tt XSPEC} notation, we used a multicolour disc
black body {\sc diskbb} \citep{mit84} plus a Comptonization model
{\sc comptt} \citep{titar94} and, when necessary, we added a
Gaussian for the Fe line with the {\sc gaussian} model and the
reflection component {\sc reflect} \citep{mag95}. This latter
component models the X-ray reflection of the comptonized radiation
from neutral or partially ionized matter, presumably the optically
thick accretion disc (Done \etal 1992, Gierli\'nski \etal 1997,
1999, Magdziarz \& Zdziarski, 1995). For this source, we always
used a fixed absorption column density~$N_{\rm
H}$~of~6~$\times$~10$^{21}~$cm$^{-2}$ \citep{bal95}. We also tied
the input soft photon temperature $kT_{\rm 0}$ of the {\sc
comptt} model to the inner disc temperature $kT_{\rm in}$ value
found by the {\sc diskbb} model. In order to compare all our data
(from epochs 1 to 4) with the same model, we show the parameters
obtained from the current fitted model described above
(multicolour disc black body plus Comptonization convolved by
reflection and Gaussian when necessary). We also tried more
complex models such as {\sc compps} and {\sc eqpair}, developed
respectively by Poutanen \& Svensson (1996) and Coppi (1999),
coupled to the {\sc gaussian} and the {\sc diskbb} (or {\sc
diskpn} for {\sc eqpair} only, see Section 3.2) models: we present
such results only for epoch 2, when the statistics were
significantly better, the instrument configurations stable and
during which the presence of a non-thermal component 
appeared more pronounced.

\begin{figure}[htbp]
\includegraphics[width=0.65\linewidth,angle=270]{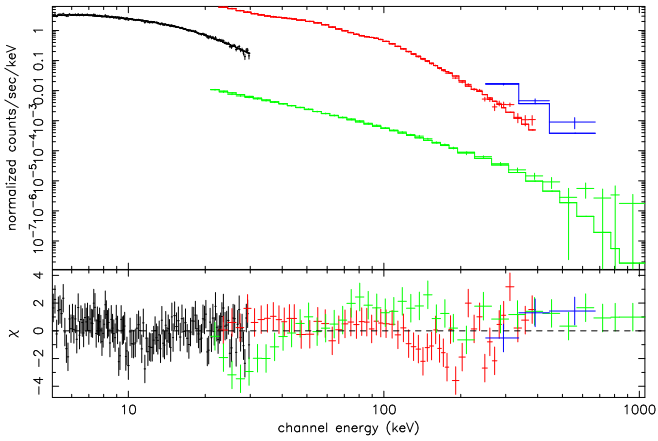}
\caption{\label{HS} Spectra of Cygnus~X-1 in 2002 December 9--11 (epoch 1) 
with the JEM-X (black), SPI (green) and IBIS (ISGRI: red; PICsIT: blue) data. 
The best-fit model is a Comptonization~model with reflection 
(see Table 2). Residuals in $\sigma$ units are also shown.}
\end{figure}

\begin{figure}[htbp]
\centering\includegraphics[width=1\linewidth]{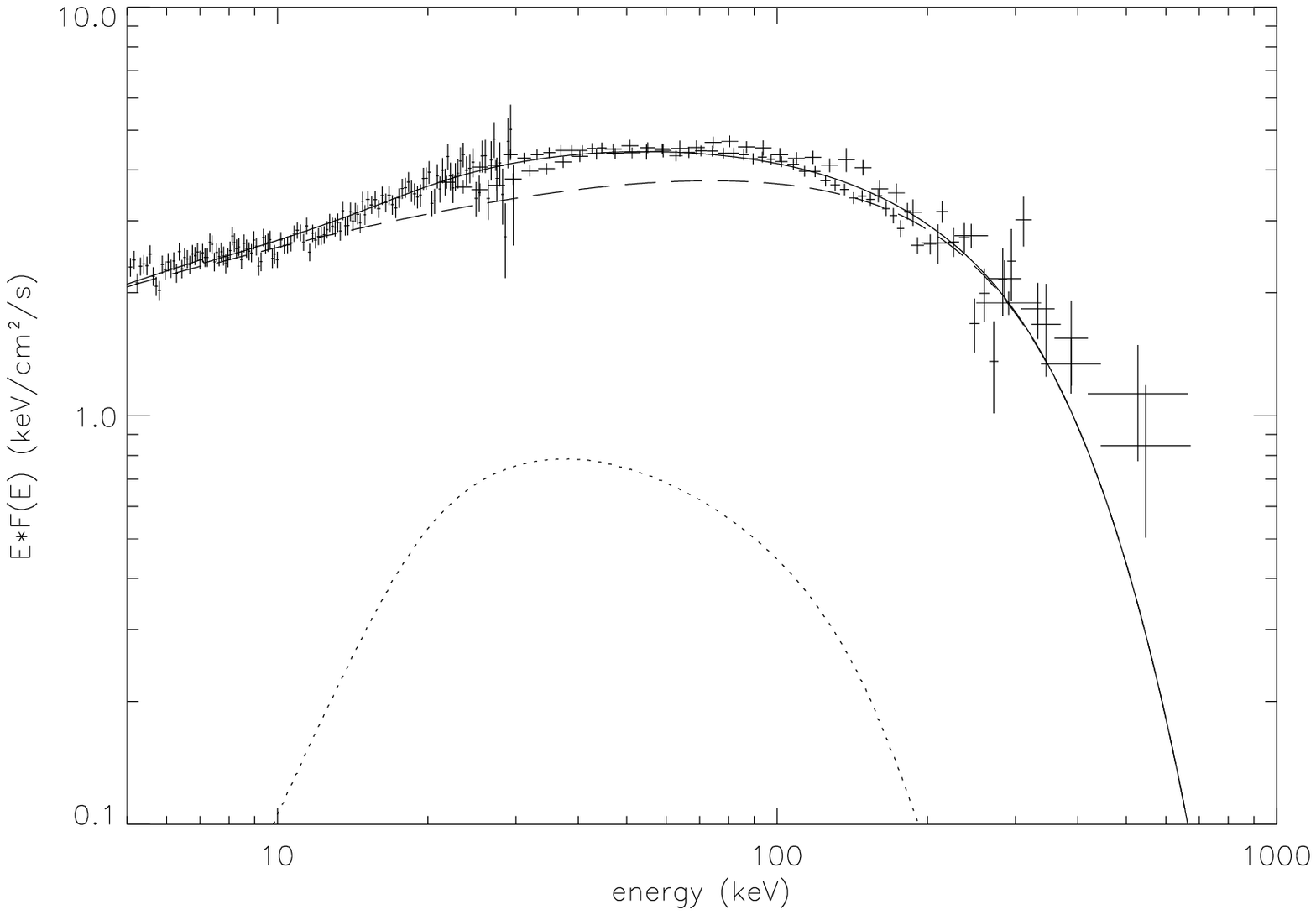}
\caption{\label{phopv} Epoch 1 unabsorbed $EF(E)$ spectrum of Cygnus~X-1 along
with the best-fit model described in Table 2 with the JEM-X, SPI and 
IBIS (ISGRI and PICsIT) data.{\it Dotted}: Reflection. {\it Long Dashes}: Comptonization.{\it Thick}: Total model.}
\end{figure}

\begin{figure}[htbp]
\includegraphics[width=0.65\linewidth,angle=270]{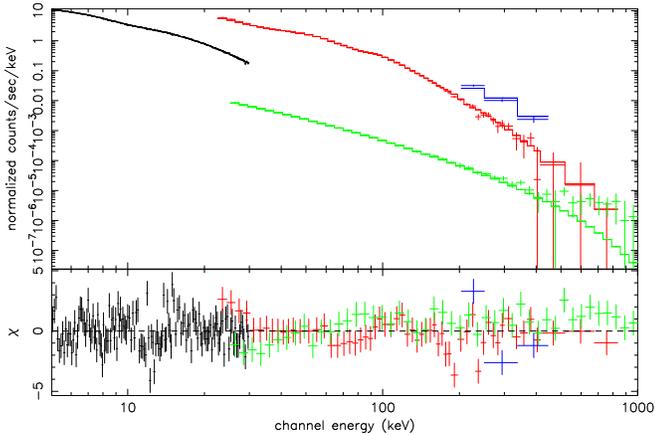}
\caption{\label{simult} Spectra of Cygnus~X-1 in 2003 June 7--11 (epoch~2)
with the JEM-X (black), SPI (green) and IBIS (ISGRI: red; PICsIT: blue) data. 
The best-fit model is a multicolour disc black body and a Comptonization model with Gaussian and reflection components (see Table 2). 
Residuals in $\sigma$ units are also shown.}
\end{figure}

\begin{table*}[htbp]
\begin{center}
\caption{\label{tab:para}~Best-fit parameters of Cygnus~X-1 for the current thermal
model in the different observation epochs.}
\begin{tabular}[h]{lllllllll}
\hline
Epoch &~~~~Dates&Disc Norm.$^a$ &$kT_{\rm in}$ or $kT_{\rm 0}$&~~$kT_{\rm e}$ &~~~~~~$\tau$ &~~~$E_{\rm Fe}$
line&~~~$\Omega/2\pi^b$&~~~~\kir \\
&~~~~(MJD)&&~~~~(keV)&~(keV) & &~~~~(keV)&&~~~(dof)\\
\hline
~~~1&52617--52620 &~~~~~~~~-&0.20 (frozen) &~~67$~_{-~6}^{+~8}$&1.98$~_{-0.23}^{+0.21}$&~~~~~~~-&~0.25$~_{-0.04}^{+0.03}$&1.45~(230)\\
~~~2&52797--52801&~~~250$~_{-59}^{+89}$&1.16~$\pm$~0.07&~100$~_{-17}^{+29}$&0.98$~_{-0.28}^{+0.25}$&7.07$~_{-0.11}^{+0.12}$&~0.57$~_{-0.06}^{+0.09}$&1.69~(236) \\
~~~3{\it{a}}&52710--52780&~~~~~~~~-&0.20 (frozen)&~~68$~_{-12}^{+22}$&2.08$~_{-0.84}^{+0.51}$&6.48~$\pm$~0.13&~0.32$~_{-0.07}^{+0.05}$&1.07~(190) \\
~~~3{\it{b}}&52801--52825&~~~312$~_{-24}^{+25}$&1.15~$\pm$~0.03&~~93~$\pm~42$&0.80$~_{-0.40}^{+0.86}$&6.40~$\pm$~0.73&~0.58$~_{-0.18}^{+0.20}$&0.93~(190)\\
~~~3{\it{c}}&~~~~52990&~~~361$~_{-67}^{+61}$&0.99~$\pm$~0.08&~~58$~_{-15}^{+54}$&1.60$~_{-0.80}^{+0.64}$&6.96~$\pm$~0.19&~0.23$~_{-0.09}^{+0.17}$&0.99~(190)\\
~~~3{\it{d}}&53101--53165&~~~~~~~~-&0.20 (frozen)&~~56$~_{-~7}^{+12}$&2.28$~_{-0.41}^{+0.30}$&6.11~$\pm$~0.26&~0.27~$\pm~$0.06&0.81~(190)\\
~~~3{\it{e}}&53240--53260&~~132~$\pm$~10&1.39~$\pm$~0.77&~~48$~_{-~6}^{+20}$&1.85$~_{-0.07}^{+0.40}$&6.49~$\pm$~0.38&~0.49$~_{-0.32}^{+0.37}$&1.56~(190)\\
~~~4&~~~~53335&~~~232$~_{-32}^{+21}$&1.16 (frozen)&~128$~_{-63}^{+84}$&0.74$~_{-0.38}^{+0.88}$&7.78$~_{-0.42}^{+0.44}$&~0.47$~_{-0.14}^{+0.18}$&0.97~(221)\\
\hline
\end{tabular}
\end{center}
Notes:\\
~a)~Disc normalization K is given by $K~=~(R/D)^{2}~\cos~\theta$
where $R$ is the inner disc radius in units of km, $D$ is
the distance to the source in units of 10 kpc and $\theta$ the
inclination angle of the disc.\\
~b)~Solid angle of the reflection component.\\
Model applied in {\tt XSPEC} notations: {\sc constant}*{\sc wabs}*({\sc diskbb}+{\sc gaussian}+{\sc reflect}*{\sc comptt}) with $N_{\rm H}$ fixed to 6~$\times$~10$^{21}$cm$^{-2}$ and $kT_{\rm 0}$ value tied to disc $kT_{\rm in}$. Errors are at 90$\%$ confidence level ($\Delta$\ki~=~2.7).
\end{table*}

\subsection{The hard state spectrum}

Figure~\ref{HS} shows the resultant count spectrum obtained
in epoch 1 (2002 December 9--11) from 5~keV up to 1~MeV
with JEM-X, SPI, IBIS/ISGRI and PICsIT data.
A simple power law model does not fit the spectra well (photon
index $\Gamma$~of~1.9~$\pm~0.1$ and reduced chi-square
\kir~=~12.90 with 213 degrees of freedom, hereafter dof).
A cutoff in the model, with a typical folding energy of
approximately 150 keV, clearly improves
the fit (\kir~=~2.12 with 212 dof) and better describes the
available data. Since a cutoff power law
is usually attributed to thermal Comptonization, we replaced
this phenomenological model by a more physical model of
Comptonization ({\sc comptt}, Titarchuk 1994).
Some residuals were still visible around
10~keV so we added a model of reflection
(with an inclination angle equal to 45°) to account for
this excess. The final tested model therefore includes
thermal Comptonization convolved by reflection ({\sc reflect}),
with solar abundances for Fe and He \citep{and89}.
We obtain a plasma temperature $kT_{\rm e}$ of 67~keV with an
optical depth $\tau$
of 1.98 and $\Omega/2\pi$~=~0.25, with \kir~=~1.45 (230~dof).
The disc black body is very weak or below the energy range of
JEM-X: this component was not used in our fits. As it gives no
contribution, we froze the $kT_{\rm 0}$ temperature of {\sc comptt}
at 0.20~keV.
Normalization constants between instruments (JEM-X, IBIS/ISGRI,
SPI, IBIS/PICsIT) are respectively equal to 1, 1.2, 1.3 and 0.9.\\
\indent Note that some residual points are visible 
in this plot, in particular for the SPI data at low energies 
(and for the IBIS data around 200 keV). This is mostly related to 
the non perfect cross-calibration between
the {\it INTEGRAL} instruments. Indeed, by fitting separately the different
instrument data, the residuals are reduced, with little change in
the spectral parameters.
Future improvement in the cross-calibration of {\it INTEGRAL} telescopes
will allow a better determination in the relative flux normalizations 
and also a better agreement of the derived spectral shapes.  
The IBIS configuration was not stable in
the first phases of the mission: in particular, the PV-Phase spectra
may suffer from the fact that the IBIS responses were built from the Crab
nebula observations performed in later periods, with a defined and
stable configuration.\\
\indent In spite of these caveats, all spectra overlap relatively
well and do define the same set of spectral parameters.
Figure~\ref{phopv} shows the resultant $EF(E)$ spectrum and its
best-fit with the JEM-X, IBIS and SPI data. Table~\ref{tab:para}
summarizes all fit results. While the 20--100~keV luminosity is
6.5~$\times$~10$^{36}$~ergs~s$^{-1}$ (with a distance to the
source fixed at 2.4~kpc), the bolometric luminosity (extrapolated
from 0.01~keV to 10~MeV) has the value of
2.2~$\times$~10$^{37}$~ergs~s$^{-1}$. The best-fit parameters we
obtain are consistent with those found in BH binaries in the HS
\citep{McClintock03} and with previous results reported on Cygnus
X-1 in the HS, in particular from the {\it INTEGRAL} data
\citep{bazz03,bouch03,pott03b}. These authors generally give
plasma temperatures between 50--100 keV and optical depths in the
1.0--2 range. In particular, our SPI spectrum is fully
compatible with the one reported by Bouchet \etal (2003). These
authors observed that the derived spectrum had an excess component
with respect to a Sunyaev \& Titarchuk (S-T) Comptonization model
(1980) and that it was better described by a cutoff power law with
$\Gamma$~=~1.5 and a cutoff energy of 155~keV. We find exactly the
same results for our SPI spectrum alone if we fit it with an S-T
model or a cutoff power law. This excess emission, with respect to
the simple S-T model, was already observed in a number of cases
for the HS spectra of BH systems, e.g., Jourdain \& Roques (1994).
However, with the Titarchuk (1994) model (which better describes
the Comptonization for high plasma temperatures), no significant
excess is observed in epoch 1 spectra of Cygnus X-1.

\subsection{The transition to a softer state in 2003 June}

\begin{figure}[htbp]
\centering\includegraphics[width=1\linewidth]{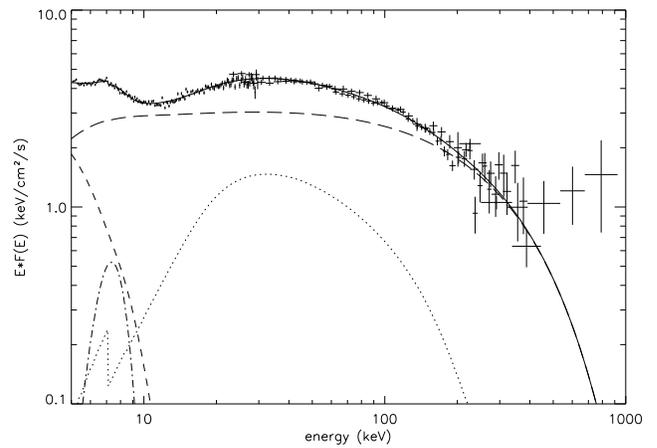}
\caption{\label{phomal} Epoch 2 unabsorbed $EF(E)$ spectrum of Cygnus~X-1 along with the best-fit model described in Table 2 with the JEM-X, SPI and IBIS (ISGRI and PICsIT) data. {\it Dashed}: Disc. {\it Dotted-Dashed}: Gaussian line.  {\it Dotted}: Reflection. {\it Long Dashes}: Comptonization. {\it Thick}: Total model.}
\end{figure}

As visible in the change of the ASM light curve of
Cygnus~X-1 (see Fig.~\ref{LCsimult}) and its corresponding
high-energy HR shown in Fig.~\ref{HR} (average value of 1.1 for
epoch 2 compared to more than 1.4 for epoch 1), the source softens
during epoch 2. Although a single power law does not fit the data
properly (\kir~=~23.51 with 245 dof), the derived slope is softer 
($\Gamma$~=~2.2~$\pm~0.1$). As for the HS spectrum analysis, 
an inspection of the residuals
reveals what kind of component can be added to build a proper
model. The need for each new component is then checked with the
results of the fits and the final model is further verified by
repeating the procedure
with components added in a different order.\\
\indent First, we included a cutoff in the power law
and we improved the $\chi^2$ by $\Delta\chi^2$~=~7.8: this
indicates that the new component is significant at more
than the 95$\%$ confidence level. A cutoff in the power law model,
with a typical folding energy of approximately 200~keV,
better describes the available data, so we
tried the physical Comptonization model ({\sc comptt}) as
we did for epoch~1.
Moreover, since very large residuals are visible in the
soft X-rays and since we suspect a transition to the IS,
we added a multicolour disc black body ({\sc diskbb}),
which is in fact required by the data
(\kir~=~14.87 and 239 dof without this component).
Some residuals around 10 keV indicate the need for
a reflection component.
Therefore, the best model for the continuum consists of a multicolour 
disc black body ({\sc diskbb}) and thermal Comptonization
({\sc comptt}) convolved by reflection ({\sc reflect},
with parameters as above). The Comptonization temperature
$kT_{\rm 0}$ was fixed to
the $kT_{\rm in}$ value returned by the {\sc diskbb}.
Finally, we added a Gaussian to account for the residuals
in the JEM-X data around 6--7 keV (\kir~=~2.42 and 237 dof
without this component).\\
\indent Table~\ref{tab:para} summarizes the best-fit parameters
and the \kir~obtained from 5~keV up to 1~MeV. We obtained a plasma
temperature $kT_{\rm e}$ of 100$~_{-17}^{+29}$~keV 
and an optical depth $\tau$ of 0.98$~_{-0.28}^{+0.25}$, 
respectively higher and lower than in epoch 1 
($kT_{\rm e}$~=~67$~_{-6}^{+8}$~keV, 
$\tau$~=~1.98$~_{-0.23}^{+0.21}$). The inner
disc temperature reached 1.16~keV and a significant line is
detected at a centroid energy of 7.07~keV, with an Equivalent
Width (EW) of 1.4~keV. With the same assumed distance, the
luminosity is 6.5~$\times$~10$^{36}$~ergs~s$^{-1}$ in the
0.5--10~keV range and 5.2~$\times$~10$^{36}$~ergs~s$^{-1}$ in the
20--100~keV band. The bolometric luminosity, extrapolated from
0.01~keV up to 10~MeV, has the value of
2.0~$\times$~10$^{37}$~ergs~s$^{-1}$; 
the disc accounts for 26~$\%$ of the total luminosity.\\
\indent The derived disc normalization is possibly
underestimated and not well constrained by the JEM-X
data (which start at 5~keV). The multicolour disc black body 
is only an approximation
of the soft component: the direct derivation of
physical parameters of the disc from the best-fit values may
suffer from some important effects \citep{merl00}.
According to these authors, the dominant effect seems
to be that, in the inner part of the disc, the opacity is
dominated by electron scattering rather than free-free absorption.
The net result is that the derived temperature given by
the $kT_{\rm in}$ parameter overestimates the
effective inner temperature by a factor of 1.7 or more \citep{shi95}.
This has an effect on the estimation of the inner disc radius as well.
Assuming an inclination angle of 45°, the derived internal
radius from the best-fit disc normalization
has the unphysical value of 4.50$~_{-0.56}^{+0.74}$~km.
For a 10~M$_\odot$ BH, this value corresponds to
0.15~R$_{\rm s}$ (where $R_{\rm s}$ is the Schwarzschild radius),
i.e., smaller than the last stable orbit
even around a maximally spinning Kerr BH.\\
\indent A comparatively strong reflection component
($\Omega/2\pi~=~$0.57$~_{-0.06}^{+0.09}$ while
it was 0.25~$_{-0.04}^{+0.03}$ for epoch 1)
is also necessary to fit the data
(\kir~=~4.57 and 237 dof without this component). Our
 $\Omega/2\pi$ value is consistent with {\it ASCA}, {\it RXTE}
\citep{gier96,gier99} and {\it BeppoSAX} observations
\citep{front01}. The best-fit model, over-plotted on data from
JEM-X, SPI, IBIS/ISGRI and PICsIT, is reported in count units in
Fig.~\ref{simult}. Figure~\ref{phomal} shows the resultant
$EF(E)$ spectrum with its best-fit. Note that we have rebinned,
for illustrative purposes, the SPI high-energy points above
750~keV to reach the level of 3$\sigma$, but the fit was performed
on the original energy channels of the spectrum (50 bins over the
22~keV--1~MeV band). Normalization constants are respectively
equal to 1, 1, 1.2 and 1. Residuals again show that improvements
in the instrument cross-calibration and responses are still needed
to obtain a fully satisfactory fit of the {\it INTEGRAL} combined
data, but we believe that the general spectral model and
parameters
are well determined.\\
\indent Considering the behaviour of the ASM light curve
(Fig.\ref{LCsimult}), the evolution of the high-energy IBIS HR
(Fig.~\ref{HR}, lower panel), the relative softness of the
spectrum and the presence of a relatively strong hard energy
emission, it appears that during the 2003 June observations
Cygnus~X-1 was in the IS (or in the FST). This is also confirmed
by radio observations of Malzac \etal (2004) who suggested that
the fluctuations of the radio luminosity were associated with a
pivoting of the high-energy spectrum. The derived
thermal Comptonization parameters are
consistent with those found in BH binaries in 
soft states \citep{McClintock03}.\\
\indent As it can be seen in Fig.~\ref{phomal}, an excess
with respect to the Comptonized spectrum above 400~keV
is observed in the SPI data that is not present in epoch 1.
We tried different models of the background,
we used different selections of pointings with two data processings
(the standard pipeline and the specific softwares developed 
by the SPI instrument team); however, this feature is always present.
To evaluate the possible presence of an instrumental
feature, we re-analyzed the Crab observation performed with
SPI on 2003 February 22--24, with the same
configuration and dither pattern as those of our Cygnus X-1 observations.
We obtained a spectrum consistent with the one reported by Roques
\etal (2003).
With 3$\%$ systematics added to the 200~ks SPI data, the Crab spectrum is
well described by a power law with $\Gamma$~=~2.1
over the 22--1000 keV energy range and no high-energy excess above 400 keV
is observed. This test shows that it is unlikely that 
the observed high-energy tail in the Cygnus~X-1 IS spectrum is 
due to a systematic effect.\\
\indent We fitted this excess with an extra component,
in addition to our current Comptonization model.
Using a power law for all the data (except JEM-X), 
we obtain a best-fit photon
index $\Gamma$ of 2.12$~_{-0.16}^{+0.31}$. The improvement
in the \kir~is relatively important (1.65 with an 
absolute chi-square reduced by 10 for 2 additional
free parameters),
making this component significant. An F-test \citep{bev92}
provides a chance probability that this improvement is due to
fluctuations of about 10$^{-2}$ (but see Protassov \etal
2002 about the limitations of the F-test).
At the same time, the parameters obtained for the current model 
are different to those obtained without the power law. 
We obtained a plasma temperature of 
50~keV, an optical depth of 2.1 and $\Omega/2\pi$ is around 0.49.
While the IBIS high-energy data do not require this additional
component, IBIS points are also consistent, due to the large error bars, with
this additional power law model.
A better \kir~(equal to 1.55, with an absolute chi-square reduced by 35, 
giving now a chance
probability of 10$^{-5}$ for the additional
power law that the improvement is due to fluctuations)
is obtained when PICsIT data are neglected.
Since the SPI telescope is certainly the best calibrated
{\it INTEGRAL} instrument at energies higher than 400-500 keV,
we consider this feature to be significant.\\
\indent Consequently, in order to account for this high-energy
emission with more physical models, we fitted the data with the
hybrid models {\sc compps} and {\sc eqpair} (coupled to the usual
disc and Gaussian line components). They combine both thermal and
non-thermal particle distributions in the calculation of the
emergent spectrum, as fully described by Poutanen \& Svensson
(1996) and Coppi (1999). Those models use a hot plasma cloud,
mainly modeled as a spherical corona around the compact object
(configuration taken in our fits) or as the well-known ``slab''
geometry, illuminated by soft thermal photons coming from an
accretion disc. These photons are Compton scattered by both
thermal (Maxwellian) and non-thermal (power law) electrons that
lose energy by Compton, Coulomb and bremsstrahlung interactions.
The number of electrons is determined by $\tau$, the corresponding
(total) vertical
Thomson optical depth of the corona.\\
\indent Both {\sc eqpair} and {\sc compps} models allow one to
inject a non-thermal electron distribution with Lorentz factors
between $\gamma_{\rm min}$ and $\gamma_{\rm max}$ and a power law
spectral index $\Gamma_{\rm p}$. In the {\sc eqpair} model, the
system is characterized by the power $L_{\rm i}$ supplied to its
different components, expressed as the dimensionless compactnesses
$l_{\rm i}~=~(L_{\rm i}\sigma_{\rm T})~/~Rm_{\rm e}c^3$ (where $R$
is the corona size). This model has two advantages with respect to
the {\sc compps}: first, it is valid both for high $kT_{\rm e}$
and low $\tau$ and vice-versa (while {\sc compps} should be used
for $\tau$~$<$~3 only). Secondly, the {\sc eqpair} model also takes 
into account the electron-positron annihilation process. We 
used the {\sc diskpn} and {\sc diksbb} routines in
the {\tt XSPEC} package to describe the soft emission from the
accretion disc in the two hybrid models ({\sc eqpair} and then
{\sc compps}). With the {\sc diskpn}, the main characteristic is
that the disc spectrum incident on the plasma is computed assuming
a pseudo-Newtonian potential for the accretion disc, extending
from 3 (the minimum stable orbit) to 500 $R_{\rm s}$.\\
\indent Both the purely thermal and the thermal/non-thermal best-fit
parameters of these hybrid models for the Cygnus X-1
spectrum of 2003 June are reported in Table~\ref{tab:eqp}.
The pure thermal models obviously do not improve the fit with
respect to the thermal Comptonization model, but when the effect
of a non-thermal electron distribution is included, the \ki~are
reduced by a significant amount and the high-energy component appears well
fitted by these models, in particular by the non-thermal
{\sc eqpair}.\\
\indent First, for {\sc eqpair}, with frozen values of $\gamma_{\rm min}$
and $\gamma_{\rm max}$
at respectively 1.5 and 1000 and with the same value for $N_{\rm H}$
and abundances as before, we obtain the dimensionless parameters
$l_{\rm h}/l_{\rm s}$~=~4.57~$_{-0.87}^{+0.04}$ and
$l_{\rm nth}/l_{\rm h}$~=~0.16~$_{-0.08}^{+0.11}$
(where $l_{\rm s}$, $l_{\rm th}$, $l_{\rm nth}$ and
$l_{\rm h}~=~l_{\rm th}+l_{\rm nth}$
correspond to the compactness in soft disc photons
irradiating the plasma, in direct thermal electron heating,
in electron acceleration and in total power supplied to
electrons in the plasma respectively).
These parameters indicate that the power in the non-thermal
component is $\sim$~16$\%$ of the total power supplied
to the electrons in the corona. With a \kir~=~1.55 (232 dof), clearly
better than the current epoch 2 model
and the pure thermal version of the {\sc eqpair} model (values 
also reported in Table~\ref{tab:eqp}{\bf )}, the derived thermal values
of $\tau$, $\Omega/2\pi$, $E_{\rm Fe}$ centroid and EW match,
within the uncertainties, the parameters obtained in
Table~\ref{tab:para}. The value of $kT_{\rm e}$ decreases from the
pure thermal model as expected and it is similar to the value we
get when adding a power law to our current model. While fitting a
soft state spectra from 0.5--200~keV with the same hybrid model,
Frontera \etal (2001) found $\Gamma_{\rm p}$ equal to
2.5~$\pm$~0.1, an EW~$\sim$~300~eV and $\Omega/2\pi$ around 0.63.
These values match well with our non-thermal results. However,
they obtained a lower $l_{\rm h}/l_{\rm s}$~$\sim$~0.36 and a
higher $l_{\rm nth}/l_{\rm h}$~$\sim$~0.77. 
For what they called a typical soft state, McConnell \etal
(2002) found a $\Gamma_{\rm p}$ compatible with ours (around 2.6)
but, again, $l_{\rm h}/l_{\rm s}$~$\sim$~0.17 and $l_{\rm
nth}/l_{\rm h}$~$\sim$~0.68 different from our values. For another
soft state, Gierli\'nski \etal (1999) found even higher values for
these latter three parameters, in the range 2.6--3.4,
0.25--1.6 and 0.77--1. All these comparisons suggest that, 
in epoch 2, Cygnus X-1 
was not in the typical soft state, but was rather in the IS.\\
\indent Second, the thermal part of the hybrid model {\sc compps}
fits the data for a \kir~=~1.72 (233 dof). 
Most of the crucial parameters like $kT_{\rm e}$, the optical
depth, $\Omega/2\pi$ and the $E_{\rm Fe}$ centroid are
compatible with those obtained with our current model (reported in
Table~\ref{tab:para}) with a comparable \kir.
Trying to improve our fits, we then used
the non-thermal part of the {\sc compps} model with an electron
distribution of index 3.72~$_{-0.22}^{+0.17}$ and we report all
the derived parameters in Table~\ref{tab:eqp}. Comparing the
different \kir~ (now equal to 1.69 for 230 dof), the improvement
is significant but not as strong as the {\sc eqpair} non-thermal
model. Again, $kT_{\rm e}$ decreases from the value obtained with
a pure thermal model, which is compatible with McConnell \etal
(2002). Using the non-thermal part of the {\sc compps} model to
fit a soft state, these authors found temperatures
($\sim$~60~keV), reflection factors and energy index $\Gamma_{\rm
p}$ ($\sim$~3.5)
again compatible with ours.\\
\indent Some authors, e.g., Belloni \etal (1996) and Ibragimov
\etal (2005, submitted) claim that the soft states observed in
1996, 1998 and 1999 were in fact IS. Modeling the spectra with
{\sc eqpair} and a non-thermal high-energy component, Ibragimov
\etal showed spectral results consistent with the non-thermal
values we report here, including a high $l_{\rm h}/l_{\rm s}$
value. According to the different soft states of Cygnus X-1
analyzed, they obtained for this parameter values between 4.2 and
12 and for $\tau$, values in the range 0.9--1.6. These results are 
consistent with ours.\\
\indent We find that
the non-thermal hybrid {\sc eqpair} model the best able to fit
our epoch 2 data: this model clearly accounts for the high-energy
tail observed. The non-thermal component represents 16$\%$ 
of the total power supplied to the electrons in the corona and 
the inferred luminosity 
in the 20-100 keV range is 6~$\times$~10$^{36}$~ergs~s$^{-1}$ 
while the bolometric one is 3.3~$\times$~10$^{37}$~ergs~s$^{-1}$.

\begin{table*}[htbp]
\begin{center}
\caption{\label{tab:eqp}~Best-fit parameters of Cygnus~X-1 for thermal/non-thermal hybrid
models in the epoch 2 observations.}
\begin{tabular}[h]{lllll}
\hline
Model & eqpair (thermal) & eqpair (non-thermal) & compps (thermal)&compps (non-thermal)\\
\hline
~~$l_{\rm h}/l_{\rm s}$&~~~~~5.3~$_{-0.4}^{+0.8}$&~~~~~~~4.57~$_{-0.87}^{+0.04}$&~~~~~~~~~~-&~~~~~~~~~~-\\
~$l_{\rm nth}/l_{\rm h}$&~~~~~~~~~~-&~~~~~~~0.16~$_{-0.08}^{+0.11}$&~~~~~~~~~~-&~~~~~~~~~~-\\
~~~~$\tau$&~~~~0.66~$_{-0.09}^{+0.02}$&~~~~~~~0.49~$_{-0.02}^{+0.24}$&~~~~0.57~$_{-0.01}^{+0.03}$&~~~1.43~$\pm$~0.10\\
~~~~$\Gamma_{\rm p}$&~~~~~~~~~~-&~~~~~~~2.4~$_{-1.0}^{+0.5}$&~~~~~~~~~~-&~~~3.72~$_{-0.22}^{+0.17}$\\
$\gamma_{\rm min}$, $\gamma_{\rm max}$&~~~~~~~~~~-&~1.5, 1000 (frozen)&~~~~~~~~~~-&~1.34, 1000 (frozen)\\
~~$\Omega/2\pi$&~~~~0.73~$_{-0.03}^{+0.07}$&~~~~~0.63~$\pm$~0.08&~~~0.60~$\pm$~0.05&~~~0.87~$_{-0.28}^{+0.12}$\\
$E_{\rm Fe}$ (keV)&~~~~7.18~$_{-0.14}^{+0.16}$~&~~~~~7.06~$\pm~0.06$&~~~7.20~$_{-0.22}^{+0.10}$&~~~7.14~$\pm$~0.09\\
EW (eV)&~~~~~~~~287&~~~~~~~~~~379&~~~~~~~~238&~~~~~~~~412\\
$kT_{\rm 0}$ (keV)&~~~~1.44~$_{-0.03}^{+0.07}$&~~~~~~~1.39~$_{-0.01}^{+0.27}$&~~~1.20 (frozen)&~~~1.20 (frozen)\\
$kT_{\rm e}$ (keV)&~~~~~~~~103$^a$&~~~~~~~~~~~42$^a$&~~~~~~~111~$_{-6}^{+3}$&~~~~~~~39~$_{-2}^{+33}$\\
\kir~(dof)&~~~~1.68 (234)&~~~~~~~1.55 (232)&~~~~~1.72 (233)&~~~~~1.69 (230)\\
\hline
\end{tabular}
\end{center}
Notes:\\
~a)~The electron temperature is calculated for the best-fit model (i.e., not free parameters).\\
Model applied in {\tt XSPEC} notations: {\sc constant}*{\sc wabs}*({\sc diskpn}+{\sc gaussian}+{\sc eqpair}) or {\sc constant}*{\sc wabs}*({\sc diskbb}+{\sc gaussian}+{\sc compps}) with $N_{\rm H}$ fixed to 6~$\times$~10$^{21}$cm$^{-2}$ and $kT_{\rm 0}$ value tied to disc $kT_{\rm in}$. Errors are at 90$\%$ confidence level ($\Delta$\ki~=~2.7).
\end{table*}

\subsection{Evolutions occurring during Epochs 3 and 4 (2003-2004)}

\begin{figure}[htbp]
\includegraphics[width=0.65\linewidth,angle=270]{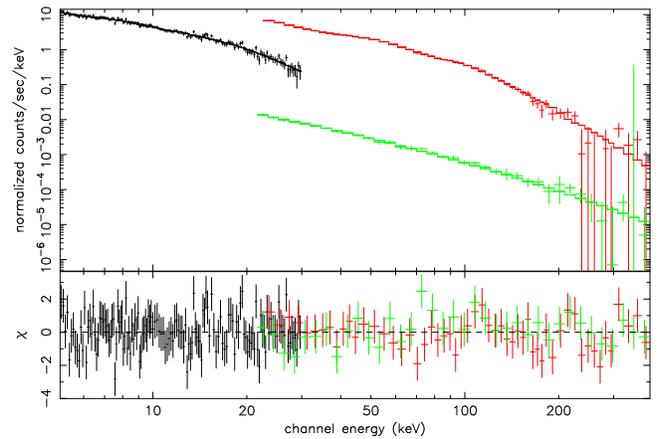}
\caption{\label{257} Epoch 4 spectra of Cygnus~X-1 with the JEM-X (black), SPI (green) and IBIS/ISGRI (red) data along with the best-fit current model 
 (see Table 2). Residuals in $\sigma$ units are also shown.}
\end{figure}

\begin{figure}[htbp]
\centering\includegraphics[width=1\linewidth]{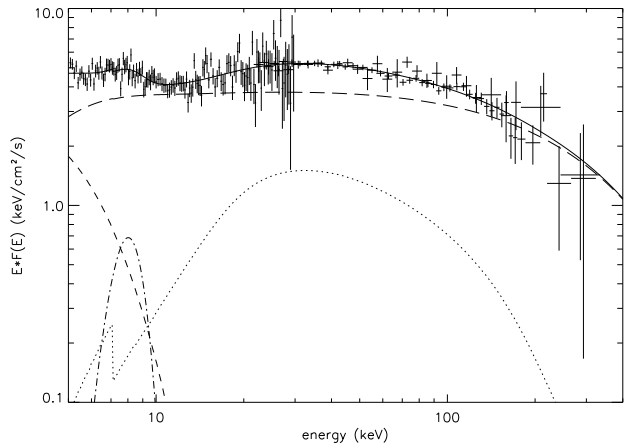}
\caption{\label{pho257} Epoch 4 unabsorbed $EF(E)$ spectrum of Cygnus~X-1 along with the best-fit model described in Table 2 with the JEM-X, IBIS/ISGRI and SPI data. {\it Dashed}: Disc. {\it Dotted-Dashed}: Gaussian line.  {\it Dotted}: Reflection. {\it Long Dashes}: Comptonization. {\it Thick}: Total model.}
\end{figure}

We separated epoch 3 in five sub-groups 
as follows:
~{\it{a)}}~before epoch 2 (MJD~$\sim$~52710--52780);
{\it{b)}} just after epoch 2 (MJD~$\sim$~52801--52825); {\it{c)}}
at MJD~=~52990 when the source switches again to a softer state;
~{\it{d)}} when a harder state is then observed
(MJD~$\sim$~53101--53165); ~{\it{e)}} observations when the ASM
count rate slightly increases while the IBIS HR decreases
(MJD~$\sim$~53240--53260). The IBIS HR shown in Fig.~\ref{HR1} 
indicates spectral evolution with time: from groups
{\it{a}} to {\it{e}}, the mean HR value is 
1.55, 0.8, 1.1, 1.4 and 1.05. Accordingly, a simple power law
model fits the data with different indexes: $\Gamma$~=~1.9, 2.3,
2.2, 1.9
and 2.3 $\pm$ 0.1 for groups {\it{a}} to {\it{e}}.\\
\indent We fitted from 5~keV to 400~keV the JEM-X and IBIS/ISGRI
spectra of the data collected during these sub-groups, using our
current Comptonization ({\sc comptt}) model for the HS and the IS
defined above. We fixed the seed photon temperature $kT_{\rm 0}$ to
the $kT_{\rm in}$ value given by the {\sc diskbb} (or frozen to
0.20~keV when the disc emission is not detected). The best-fit
parameters are listed in Table~\ref{tab:para}. Normalization
constants between the two instruments are very close to each other
($\sim$~1--1.2 for IBIS/ISGRI when JEM-X constant is frozen to 1).
Variations in amplitude and overall spectral shape are observed
between the epoch 3 sub-groups. While no disc emission is detected
for groups {\it a} and {\it d}, a disc component is required for
groups {\it b}, {\it c} and {\it e}. The EWs are also changing: the
softer the source, the larger the EW ($\sim$~110~eV for group
{\it{a}} compared to $\sim$~780~eV for group {\it{e}}).
These results and the IBIS HR (Fig.~\ref{HR1}) indicate that, 
during sub-groups 3 {\it a} and {\it d}, Cygnus~X-1 was in a HS
(as in epoch 1) while, in sub-groups 3 {\it b}, {\it c} and {\it e}, 
the source was in a softer state.\\
\indent During epoch 4, the simple power law slope is
$\Gamma$~=~2.2~$\pm$~0.1 and the mean IBIS HR value is 1.1. This
indicates that the source was again in a softer state than the HS.
The count spectrum is presented in Fig.~\ref{257}.
Figure~\ref{pho257} shows the resultant $EF(E)$ spectrum and its
best-fit model overplotted on the data obtained from JEM-X,
IBIS/ISGRI and SPI (high-energy IBIS points are rebinned at the
3$\sigma$ level above 200~keV for illustrative purposes).
Cross-calibration constants are respectively 1, 1 and 
1.3. Using the current Comptonization ({\sc comptt}) model defined above
for epoch 2, we present in Table~\ref{tab:para} the best-fit
parameters obtained. Since we obtained a best-fit solution 
similar to the one of epoch 2,
we fixed, to define the errors for the other parameters, 
the $kT_{\rm in}$ to
the epoch 2 value of 1.16 keV and we obtained
$kT_{\rm e}$~=~128~keV, $\tau$~=~0.74,
$\Omega/2\pi$~=~0.47~$_{-0.14}^{+0.18}$
(close to the epoch 2 value of
0.57~$_{-0.06}^{+0.09}$) and
$E_{\rm Fe}$~=~7.78~keV. 
Althought the temperature is little constrained 
and errors are large, these values indicate that the source 
was again in a rather soft state. 

\section{Discussion}

During the broad-band (5~keV--1~MeV) {\it INTEGRAL} observations
of Cygnus X-1 presented in this paper,
the source was detected in at least two different spectral
states.
For epoch 1 (part of PV-Phase) and during some of the GPS observations
(epoch 3, sub-groups {\it {a}} and {\it {d}}), the source was
in the typical HS with a high-energy spectrum extending
up to 800~keV, well characterized by a thermal Comptonization model.
Parameters are typical of BH binaries and consistent with previous
observations of Cygnus X-1 in the HS.
We find that the Comptonization component must be modified by
reflection. The subtended angle we derived is also 
compatible with values previously found for this source in the HS, 
either from {\it Ginga} and {\it CGRO}/OSSE 
($kT_{\rm e}\sim$~100~keV, $\tau~\sim~$1--2,
and $\Omega/2\pi$~$\sim$~0.19--0.34, Gierli\'nski \etal 1997), 
or from {\it BeppoSAX}
($kT_{\rm e}\sim$~60~keV, $\tau~\sim~$1--2 and 
$\Omega/2\pi~=~$0.25, Frontera \etal 2001).
Di Salvo \etal (2001) observed even higher temperatures for the
HS, from 111 to 140~keV,
and $\Omega/2\pi~\sim$~0.1--0.3, within our error bars.
Also, our results are compatible with those reported
by Bazzano \etal (2003), Bouchet \etal (2003)
and Pottschmidt \etal (2003b)
who used {\it INTEGRAL} data from other
PV-Phase observations, when Cygnus X-1
was also in the HS. We did not detect any Fe line nor disc emission
during this period. Indeed, we could determine a 90$\%$ 
confidence level 
upper limit of 94~eV EW for a narrow line ($\sigma$~=~0.1~keV) 
at 6.7~keV and of 172~eV EW for a broad one ($\sigma$~=~1~keV).
These limits are compatible with the typical EWs 
($\sim$~150~eV) of broad lines observed in HS from this 
BH system \citep{front01}. The lack of strong Fe line and disc 
emission is not surprising
since, in the HS, the disc does not extend close to the BH, 
its inner disc temperature is low and 
contribution at~$>$~5~keV is negligible.\\
\indent Cygnus~X-1 underwent a clear evolution to a softer
state from epoch 1 to epoch 2. We indeed
observed in 2003 June {\it INTEGRAL} data hardness
variations (Fig.~\ref{HR1} and \ref{HR}) and photon index
changes, along with the appearance of a significant soft
component well fitted by a disc black body model.
The changes in the high-energy component are obvious from
the best-fit parameters
of the current model reported in Table~\ref{tab:para}.
The Comptonization parameter $y$ is characterized by both
values ($kT_{\rm e}$,~$\tau$)
where $y=(kT_{\rm e}/m_{\rm e} c^2)~\max~(\tau,\tau^2)$.
While $y$~$\sim$~0.51$~_{-0.09}^{+0.10}$ for epoch 1, its
value drops to 0.19$~_{-0.06}^{+0.07}$ for epoch 2.\\
\indent As shown previously, these changes were correlated with 
evolutions of the {\it RXTE}/ASM light curve (Fig.~\ref{LCsimult})
which indicate the rise of the soft disc emission. The increase in
disc emission (up to about 26$\%$ of the bolometric luminosity)
combined with an increase of the inner disc temperature to about 
1.16~keV suggest that the accretion disc has extended down very
close to the BH horizon. However, the inferred inner radius (4.5~km) 
is not compatible with the size of the innermost 
stable circular orbit, even for a maximally spinning Kerr BH.
Besides uncertainties on the inclination angle and on the disc
normalization, a number of effects can lead to such unphysical
estimations. As mentioned in Section 3.2, electron scatterings 
can produce a diluted black body \citep{shi95} and cause the observed
temperature $kT_{\rm in}$ to be higher and the disc normalization lower. 
Consequently, the inner disc radii could be 
underestimated by a factor of 5 or more
\citep{merl00}. Another possible explanation for such a high
temperature could be that the disc emission is Comptonized by a
different warm corona at a temperature of a few keV, as sometimes 
suggested \citep{front01,mal05}.\\
\indent The softening of the hard component and the appearance of
the disc emission were accompanied by significant increase in
reflection and changes in Fe line emission. During epoch 2 data,
the Fe line is needed in the spectra (while not for epoch 1) and
$\Omega/2\pi$ increases (it is more than twice the value 
found for epoch 1):
this again indicates that a larger disc, closer to the BH,
reflects more radiation coming from the hot plasma.
Figures~\ref{phopv} and~\ref{phomal} show the resultant unabsorbed
$EF(E)$ spectra for both epochs 1 and 2: we clearly see
differences in amplitudes and in the relative contributions of the
various components, from soft to hard X-rays/$\gamma$-rays.
Similar spectral transitions are observed during epochs 3 and 4:
$y$~=~0.57$~_{-0.34}^{+0.27}$ for epoch 3 sub-group {\it{a}}
(while it drops to 0.14$~_{-0.09}^{+0.16}$ for group {\it{b}}),
and $y$~=~0.18$~_{-0.13}^{+0.24}$ for epoch 4.
The $y$ evolution, combined with the changes in the IBIS HR and in 
the ASM light curve, indeed shows transitions from the HS 
(epoch 1, epochs 3 {\it{a}} and {\it{d}}) to softer states 
(epochs 2, 3 {\it{b}}, {\it{c}}, {\it{e}} and 4).\\
\indent The transition to a soft state is generally attributed to
an increase in the accretion rate. In this case, the total
luminosity should increase significantly. For example, the total
luminosity measured by Frontera \etal (2001) in the TDS of
Cygnus~X-1 was about a factor of 3 higher than in the HS, with a
contribution of the disc emission higher than 50$\%$ of the total
luminosity. From our analysis, the spectral transition seems to
occur with only a slight change in the bolometric luminosity,
mainly due to the appearance of a high-energy tail which 
represents less than 2$\%$ of the Eddington luminosity
($\sim$~1.5~$\times$~10$^{39}$~ergs s$^{-1}$ for a 10~M$_\odot$
BH). Therefore, we can interpret the component evolution as a
spectral pivoting between soft and hard X-rays at almost constant
luminosity. This has already been reported for other BH such as
for the microquasar XTE~J1550-564 \citep{rod03}.
When the total luminosity does not change
much, the spectral evolution cannot be explained simply by a
large variation in the accretion rate. An additional physical
parameter (for example linked to the temperature and geometry of
the Comptonizing cloud or to the magnetic field) must probably
vary in the system to trigger the spectral changes from
the HS to a softer state at almost constant luminosity. \\
\indent While our data start at 5~keV, leading to a possible
underestimation of the bolometric luminosity
(which in a real soft state is dominated by the soft component),
the near constancy of the luminosity between epochs 1 and 2 and
other facts show that the soft state we observed in this latter period is
not the typical TDS observed in Cygnus~X-1. 
Our spectral analysis and the comparisons with previous results 
obtained on Cygnus~X-1 (see Section 3.2) led us to conclude that, 
during epoch 2, the source was rather in the IS.
The need for a cutoff in the high-energy spectrum
of epoch 2 shows that the thermal Comptonization is still a
dominant process for the high-energy component.
The hybrid thermal/non-thermal models
lead to a significant improvement
in the fit of our data, compared to the thermal Comptonization models,
like {\sc comptt}, or the pure thermal versions of 
{\sc eqpair} and {\sc compps}.
However, our best-fit non-thermal parameters
are sometimes intermediate between hard and soft state values
(like $l_{\rm h}/l_{\rm s}$, $l_{\rm nth}/l_{\rm h}$ or $\Gamma_{\rm p}$):
this again shows that Cygnus~X-1 was in the IS (or in the FST).
Such a conclusion was also reached by Malzac \etal (2004) 
on the basis of the observed correlation of the spectral hardness 
with the radio flux during this period.\\
\indent In recent years, it has become apparent that in the
HS, the BH binaries become bright in radio and display clear
correlations between the X-ray and radio luminosities 
\citep{corb03,gallo03} as often observed for Cygnus~X-1 
\citep{brock99,gleiss04b,nowak05}. Models
where the base of a compact jet plays a major role in
the physical processes of such BH systems have been proposed (Markoff
\etal 2001). In such scenarios, the high-energy emission seen
during the HS is interpreted as synchrotron emission from the jet
that extends from radio to hard X-rays, naturally explaining the
correlations observed during the HS. Those models explain the
observed (or inferred) outflows: the radio emission is
proportional to the jet power which in turn is correlated with the
accretion rate and with the X-ray emission \citep{heinz03,merl03}
as discussed for GX~339-4 \citep{zd04b}. However, detection
of radio emission was reported during the TDS, for example in the
BH XTE~J1650-500 \citep{corb04} and the scenario
increases in complexity (see Fender \etal 2005 for a complete
review). 
More recently, Markoff \etal (2004) have proposed jet 
models where the synchro-self Compton or the external Comptonization 
radiation are the dominant processes generating 
X-ray spectra in BH binaries. These models seem to fit the experimental
data as well as the thermal Comptonization models do.\\
\indent While we did not observe any significant high-energy tail
during the PV-Phase observations, it appears that, in the SPI
spectra of 2003 June, when the source was in the IS, 
some data points above 400~keV are not well
described by the thermal Comptonization models (see
Fig.~\ref{simult} and \ref{phomal}). All the
tests we have performed using different SPI background models and
the Crab nebula spectrum indicate that this feature comes from
the source. Similar excess over a Comptonization law has been
previously observed, in particular with {\it CGRO}/OSSE
\citep{mc02} which detected a power law like component extending
beyond 1~MeV during a (so classified) TDS of Cygnus~X-1.
In our case, the SPI data of the observed IS spectrum of Cygnus~X-1
imply a rather bright high-energy tail.
The bolometric luminosity of the source, including this new component,
is 1.65 times higher than the one obtained
with the best-fit thermal current model.\\
\indent This kind of steep power law, without a high-energy break 
(at least up to 1~MeV)
was modeled in the past including a non-thermal component in the
accretion flow, with the so-called hybrid thermal/non-thermal
models from Poutanen \& Svensson (1996) and Coppi (1999). Using
these same models, we obtain a better fit of our data, which
indicate the necessity to include a non-thermal distribution of
electrons, with a power law energy index between 2.4--3.7, depending
on the model used. $\Gamma_{\rm p}$ between 2 and 3 are expected
from shock acceleration models. We find that most of the crucial
parameters such as $\tau$, $\Omega/2\pi$, $E_{\rm Fe}$ centroid,
EW or $kT_{\rm e}$ are compatible with previous results reported
by McConnell \etal (2002), Frontera \etal (2001) and Gierli\'nski
\etal (1999). With the non-thermal {\sc eqpair} model, we found an
unabsorbed bolometric luminosity of
3.3~$\times$~10$^{37}$~ergs~s$^{-1}$ (higher than the one obtained
with the current thermal model). This value is $\sim$~1.9 times
lower than the one observed by McConnell \etal (2002) in the (so
classified) TDS: this, along with the differences seen 
in the $l_{\rm h}/l_{\rm s}$, $l_{\rm nth}/l_{\rm h}$ and
$\Gamma_{\rm p}$ values (see Section 3.2), definitely shows 
that our epoch 2 observations did not
happen during a TDS but rather in the IS, with an intermediate luminosity.\\
\indent Alternatively, Comptonization on a population of
(thermalized) electrons with bulk motion, e.g., Titarchuk \etal
(1997), Laurent \& Titarchuk (1999), is sometimes invoked to
explain the power law high-energy component. This could describe
the observed high-energy emission seen in epoch 2, as it predicts
$\Gamma>2$ or even softer, even if the lower energy
cutoff should be around 100~keV. In addition to hybrid thermal/non-thermal
models presented above, a stratified Comptonization region,
providing a larger range of both electron temperatures 
and optical depths,
could model the spectrum. Ling \etal (1997) reached the same
conclusions based on Monte Carlo modeling of {\it CGRO}/BATSE
spectra combined with (non contemporaneous) {\it CGRO}/COMPTEL
data (McConnell \etal 1994). Thermal gradients are incorporated
into several other models, which then lead to the generation of a
high-energy tail, e.g., Skibo \& Dermer (1995),
Chakrabarti \& Titarchuk (1995) and Misra \& Melia (1996).\\

\section{Summary and conclusions}

Using the broad-band capability of {\it INTEGRAL}, it
has been possible to accumulate a large amount of simultaneous
data on Cygnus~X-1 between 5~keV--1~MeV and to follow its spectral
evolution from 2002 November to 2004 November. These data were
helpful to characterize the evolution of the Comptonization
parameters of the source correlated to the presence of a
variable disc emission, indicating transitions between the HS and
a softer (Intermediate) state. We also observed the
presence in the SPI data of a high-energy tail during the IS (or the FST), 
emerging from the Comptonization component between 400~keV--1~MeV 
and probably associated with a non-thermal component. 
The extent to which the spectrum hardens at energies approaching 1~MeV
has now become an important issue for theoretical modeling of the
accretion processes and radiation mechanisms in BH binaries. Data
from both IBIS and SPI instruments offer the best opportunity to
define more precisely the high-energy X-ray binary spectra.
We hope to further investigate this using {\it INTEGRAL} data 
from this source and other bright BH X-ray binaries.

\section*{Acknowledgments}
We thank the anonymous referee for his/her helpful comments and
suggestions. MCB thanks J. Paul for a careful reading and comments
on the manuscript. AAZ has been supported by KBN grants
PBZ-KBN-054/P03/2001, 1P03D01827 and 4T12E04727. We thank the ESA
ISOC and MOC teams for they support in scheduling and operating
observations of Cygnus~X-1. The present work is based on
observations with {\it INTEGRAL}, an ESA project with instruments
and science data center funded by ESA member states (especially
the PI countries: Denmark, France, Germany, Italy, Switzerland,
Spain, Czech Republic and Poland, and with the participation of
Russia and the USA).

\end{document}